\documentclass[reprint,superscriptaddress,aps,twocolumn,showpacs]{revtex4-1}
\usepackage{amssymb}
\usepackage{amsmath}
\usepackage{graphicx}
\usepackage{multirow}
\usepackage{hyperref}
\usepackage{xcolor}
\usepackage{tabularx}
\allowdisplaybreaks

\begin{document}
\title{Study of $1^{--}$ P-wave charmoniumlike and bottomoniumlike tetraquark spectroscopy}

\author{Zheng Zhao}
\email[]{zhaozheng1022@hotmail.com}
\author{Attaphon Kaewsnod}
\author{Kai Xu}
\author{Nattapat Tagsinsit}
\affiliation{School of Physics and Center of Excellence in High Energy Physics and Astrophysics, Suranaree University of Technology, Nakhon Ratchasima 30000, Thailand}
\author{Xuyang Liu}
\affiliation{School of Physics and Center of Excellence in High Energy Physics and Astrophysics, Suranaree University of Technology, Nakhon Ratchasima 30000, Thailand}
\affiliation{School of Physics, Liaoning University, Shenyang 110036, China}
\author{Ayut Limphirat}
\author{Yupeng Yan}
\email[]{yupeng@sut.ac.th}
\affiliation{School of Physics and Center of Excellence in High Energy Physics and Astrophysics, Suranaree University of Technology, Nakhon Ratchasima 30000, Thailand}

\date{\today}

\begin{abstract}
\indent The masses of $1^{--}$ P-wave charmonium-like and bottomonium-like tetraquark states are calculated in a constituent quark model (CQM) where the Cornell-like potential and Breit-Fermi interaction are employed. All model parameters were imported from previous work, and predetermined by studying the low-lying conventional S- and P-wave light, charmed, bottom, charmonium, and bottomonium meson mass spectra. 
{The lowest $1^{--}$ tetraquark mass is predicted to be around 4.15 GeV.}
The decay widths of $1^{--}$ P-wave tetraquark states are calculated for possible two-body strong decay channels within the rearrangement mechanism, including $\omega \chi_{cJ}$ and $\eta J/\psi$ for charmonium-like tetraquarks, and $\omega \chi_{bJ}$ for bottomonium-like tetraquarks.
The theoretical results are compared with the selected exotic states, also known as Y states, and tentative assignments are suggested. 
This study suggests that $\psi(4230)$, $\psi(4360)$, $\psi(4660)$, and $\Upsilon$(10753) may be P-wave tetraquark states and that multiple states might exist around 4.36 GeV. 
\end{abstract}

\maketitle

\section{Introduction}\label{sec:Int}
The study of exotic hadrons has gained significant attention in recent years, particularly in the context of the quark model. While the conventional meson and baryon spectra are well described by quantum chromodynamics (QCD) in terms of quark-antiquark and three-quark configurations, respectively, experimental discoveries of unconventional states challenge this simple classification. Among these exotic candidates, compact tetraquarks, bound states of two quarks and two antiquarks, offer a compelling framework to explain certain resonances that do not fit within the conventional meson spectrum~\cite{Liu:2019zoy,Chen:2022asf}.

Charmonium-like tetraquarks, which contain a charm and anticharm quark pair along with an additional light quark pair, are particularly intriguing due to their potential connection to the so-called Y states observed in electron-positron annihilation experiments. In the latest Particle Data Group (PDG)~\cite{PDG} naming scheme, the Y states have been renamed as $\psi$ states because they share the same quantum numbers $(J^{PC} = 1^{--})$ as the $J/\psi$ family.

The Y(4230), an exotic state confirmed by multiple experimental collaborations and considered part of the broader family of charmonium-like states, was first detected by the BaBar Collaboration in the $e^+e^- \to\pi^+\pi^-J/\psi$ channel~\cite{BaBar:2005hhc}. The Y(4360) was later observed by the Belle Collaboration in the $e^+e^- \to \gamma \pi^+ \pi^- \psi(2S)$ process~\cite{Belle:2007umv}. Subsequently, the Y(4660) was discovered by BaBar in the initial state radiation process $e^+e^- \to \gamma_{\text{ISR}} \pi^+ \pi^- \psi(2S)$~\cite{BaBar:2012hpr}. 
More recently, the Y(4484) was observed by the BESIII collaboration in the $K^+K^-J/\psi$ channel with a statistical significance greater than 8$\sigma$~\cite{BESIII:2022joj}. Additionally, the Y(4544) was reported by BESIII in the $\omega\chi_{c1}$ channel~\cite{BESIII:2024jzg} with a significance of 5.8$\sigma$.  Notably, the mass of the Y(4544) is significantly higher than that of the Y(4484), suggesting possible differences in their underlying structure.

The $\Upsilon(10753)$ bottomonium-like state was observed by the Belle Collaboration in the $e^+e^- \to \pi^+\pi^- \Upsilon(nS)$ process~\cite{Belle:2019cbt}. Recently, the Belle II Collaboration reported another observation that strongly supports the existence of the radiative transition process $\Upsilon(10753) \to \omega\chi_{bJ}(1P)$~\cite{Belle-II:2022xdi}. The energy dependence of the Born cross sections for $\Upsilon(10753)$ in the $\omega\chi_{bJ}(1P)$ channel~\cite{Belle-II:2022xdi} was found to be consistent with the shape of the $\Upsilon(10753)$ state in the $\pi^+\pi^- \Upsilon(nS)$ channel~\cite{Belle:2019cbt}. Belle-II concluded that the internal structure of $\Upsilon(10753)$ may differ from that of $\Upsilon(10860)$, where the latter is well understood as a predominantly conventional bottomonium state. 

The exotic states exhibit properties that cannot be easily accommodated with conventional quarkonium expectations, suggesting the need for alternative interpretations. 
Various theoretical pictures for charmonium-like exotic states, including charmonium core admixed with coupled channels~\cite{Lu:2017yhl,Li:2009zu,Shah:2012js,Wang:2019mhs,Wang:2022jxj}, charmonium hybrids~\cite{HadronSpectrum:2012gic,Luo:2005zg,Zhu:2005hp,Oncala:2017hop,Berwein:2015vca}, hidden-charm molecular states~\cite{Chiu:2005ey,Qin:2016spb,Ding:2008gr,Close:2009ag,Close:2010wq}, and hidden-charm tetraquark states~\cite{Maiani:2005pe,Ali:2017wsf,Drenska:2009cd,Ebert:2008kb,Wang:2024ciy,Wang:2024qqa,Wang:2018ejf}, have been proposed to explore their internal structure and mass spectra, providing valuable insights into their nature. 

{Since the inception of the quark model~\cite{Gell-Mann:1964ewy}, the tetraquark picture has been widely employed and further developed to understand the internal structure of exotic states~\cite{Jaffe:1976ig,Jaffe:1976ih}, not only in charmonium-like systems. For exotic states in the four-heavy-quark region, one may refer to the fully-heavy tetraquark picture~\cite{Bai:2024flh,Wu:2024ocq,Mistry:2024zna,Liu:2024pio,Bai:2024ezn,Wu:2024tif,Chen:2024bpz,Wu:2024euj,Zhang:2023ffe,Yu:2022lak,Zhang:2022qtp,Mutuk:2022nkw,Wang:2022yes,Chen:2022sbf,Hu:2022zdh,Liu:2021rtn,Wang:2021mma,Wang:2021kfv}. Similarly, for light exotic states, the fully-light tetraquark picture has also been considered~\cite{Zhao:2021jss,Agaev:2018fvz,Ebert:2008id,Wang:2024pgy,Xin:2022qnv,Santopinto:2006my,Lodha:2024bwn}.}

In the previous work, the Y(4230) and the $\Upsilon(10753)$ could not be accommodated within a conventional heavy quarkonium picture including $S-D$ mixing~\cite{Zhao:2023hxc}. In this work, we focus on studying the mass spectrum and decay properties of the $1^{--}$ P-wave charmonium-like and bottomonium-like compact tetraquark states (which we refer to as $Y_c$ and $Y_b$ for convenience) within a constituent quark model framework.

The paper is organized as follows. In Sec.~\ref{sec:TM}, all the possible configurations of color, spin, and spatial degrees of freedom of tetraquark states are introduced, and a constituent quark model developed from our previous work \cite{Zhao:2020jvl, Zhao:2021jss} is briefly reviewed. 
In Sec.~\ref{sec:RND}, $1^{--}$ P-wave tetraquark mass spectra and decay branching ratios are evaluated in the constituent quark model. The theoretical results are compared with experimental data of Y states and tentative assignments for $1^{--}$ tetraquarks are suggested.
A summary is given in Sec.~\ref{sec:SUM}. The details of the tetraquark spatial wave function, and cross terms between different configurations from one-gluon exchange interaction are shown in the Appendix.

\section{\label{sec:TM}THEORETICAL MODEL}

\subsection{Quark configurations of tetraquark}

The construction of tetraquark states follows the principle that a tetraquark state must be a color singlet, meaning the color wave function of the tetraquark must be a $[222]_1$ singlet within the $SU_c(3)$ group. In this work we consider only charmonium-like and bottomonium-like ($q_1Q_2\bar q_3 \bar Q_4$) states, where $q_1$ and $Q_2$ are a light quark and a heavy quark, and $\bar q_3$ and $\bar Q_4$ are a light antiquark and a heavy antiquark, respectively. 

The permutation symmetry of the two-quark cluster ($qQ$) in tetraquark states is described by the Young tableaux $[2]_6$ and $[11]_{\bar 3}$ of the $SU_c(3)$ group, while the color configuration of the two-antiquark cluster ($\bar q\bar Q$) consists of a $[211]_3$ triplet and a $[22]_{\bar 6}$ antisextet. 

Thus, a $[222]_1$ color singlet for tetraquark states demands the following configurations: $[2]_6(q_1Q_2)\otimes[22]_{\bar 6}(\bar q_3\bar Q_4)$ and $[11]_{\bar 3}(q_1Q_2)\otimes[211]_3(\bar q_3\bar Q_4)$, which correspond to the color sextet-antisextet ($6_c \otimes \bar 6_c$) and triplet-antitriplet ($\bar 3_c \otimes 3_c$) configurations, respectively. The explicit color wave function for each tetraquark color configuration can be found in the previous work~\cite{Zhao:2020jvl}.

The $J^{PC}=1^{--}$ Y states can be described by $L=1$ tetraquark configurations with $S=0$ or $S=2$. Thus, two possible spin combinations for tetraquarks are considered as follows: $\left[\psi_{[s=1]}^{qQ}\otimes\psi_{[s=1]}^{\bar q \bar Q}\right]_{S=0,2}$, and $\psi_{[s=0]}^{qQ}\otimes\psi_{[s=0]}^{\bar q \bar Q}$. 

The complete basis is constructed by coupling the harmonic-oscillator (HO) wave functions, taking the general form,
\begin{eqnarray}\label{eqn::spatial}
\psi_{NL} &=
\sum_{{n_{\chi_i},l_{\chi_i}}}
 A(n_{\chi_1},n_{\chi_2},n_{\chi_3},l_{\chi_1},l_{\chi_2},l_{\chi_3}) \nonumber \\
& \times \psi_{n_{\chi_1}l_{\chi_1}}(\vec \chi_1\,) \otimes\psi_{n_{\chi_2}l_{\chi_2}}(\vec \chi_2\,)\otimes\psi_{n_{\chi_3}l_{\chi_3}}(\vec \chi_3),
\end{eqnarray}
where $\psi_{n_{\chi_i}l_{\chi_i}}$ are HO wave functions. The sum ${n_{\chi_i},l_{\chi_i}}$ is over $n_{\chi_1},n_{\chi_2},n_{\chi_3}, l_{\chi_1},l_{\chi_2},l_{\chi_3}$. The relative Jacobi coordinates $\vec \chi_1$, $\vec \chi_2$ and $\vec \chi_3$ are defined as
\begin{flalign}\label{eqn::JC}
&\vec \chi_1=\frac{1}{\sqrt 2}(\vec r_1-\vec r_2), \nonumber \\
&\vec \chi_2=\frac{1}{\sqrt 2}(\vec r_3-\vec r_4), \nonumber \\
&\vec \chi_3=\frac{m_u\vec r_1+m_Q\vec r_2}{m_u+m_Q}-\frac{m_u\vec r_3+m_Q\vec r_4}{m_u+m_Q}, 
\end{flalign}
where $\vec{r}_{j}$ denotes the coordinate of the jth quark. The reduced quark masses are defined as
\begin{flalign}\label{eqn::rqm}
&u_1=\frac{2m_um_Q}{m_u+m_Q}, \nonumber \\
&u_2=\frac{2m_um_Q}{m_u+m_Q}, \nonumber \\
&u_3=\frac{m_u+m_Q}{2},
\end{flalign}
$m_u$ and $m_Q$ are the masses of the light and heavy quarks, respectively. $N$ and $L$ are respectively the total principal quantum number and total angular momentum, and $L=l_{\chi_1}+l_{\chi_2}+l_{\chi_3}=1$ for a P-wave tetraquark and $N= (2n_{\chi_1}+ l_{\chi_1})+(2n_{\chi_2}+ l_{\chi_2})+(2n_{\chi_3}+l_{\chi_3})$. 

{We stress that the Jacobi coordinates in Eq.~(\ref{eqn::JC}) are introduced for convenience when constructing the complete HO basis and evaluating matrix elements, and do not impose a dynamical assumption on the internal structure of the tetraquark. Other commonly used coordinate sets, often referred to as H- and K-type, are related to Eq.~(\ref{eqn::JC}) by linear transformations of the same four-body relative coordinates. Likewise, different color coupling schemes associated with these coordinate choices are connected by basis transformations (e.g., $\bar 3_c\!-\!3_c$ and $6_c\!-\!\bar 6_c$ can be written as linear combinations of $1_c\!-\!1_c$ and $8_c\!-\!8_c$). Therefore, in a sufficiently large basis, the physical observables are independent of the chosen coordinate representation~\cite{Yang:2020fou}.}

\subsection{Hamiltonian}

{ We study meson and tetraquark systems within the non-relativistic quark model~\cite{Godfrey:1985xj,Schoberl:1986bv,Barnes:2005pb}. The Hamiltonian has been widely employed in recent years~\cite{Wang:2021kfv,Wu:2024ocq} and takes the form,
\begin{flalign}\label{eqn::ham}
H= &H_0+ H_{so}, \nonumber \\
H_{0} = &\sum_{k=1}^{N} (\frac12M^{ave}_{k}+\frac{p_k^2}{2m_{k}})+\sum_{i<j}^{N}(-\frac{3}{16}\vec\lambda_{i}\cdot\vec\lambda_{j})V_{cen}(r_{ij}),  \nonumber \\
H_{so} = &\sum_{i<j}^{N}(-\frac{3}{16}\vec\lambda_{i}\cdot\vec\lambda_{j})(V_{so}(r_{ij})), 
\end{flalign}
where $M^{ave}_k$ denotes the spin-averaged mass as $\frac{1}{4}M_{PS}+\frac{3}{4}M_V$. $m_k$ are the constituent quark masses. $\vec \lambda_{i}$ is the quark color operator in SU(3). The Cornell-like potential and one-gluon-exchange potentials are employed as the central potential, taking the form,
\begin{flalign}\label{cornell}
&V_{cen}(r_{ij})=A_{ij} r_{ij}-\frac{B_{ij}}{r_{ij}}+V_{ss}(r_{ij}), 
\end{flalign}}
where the spin-spin interaction $V_{ss}(r_{ij})$ originates from the one-gluon-exchange potential and takes the form:
\begin{flalign}\label{Hsd}
&V_{ss}=\frac{1}{6m_im_j}\Delta V_V(r)\vec\sigma_{i}\cdot\vec\sigma_{j}=\frac{2B_{ij}\sigma_{ij}^3}{3m_im_j\sqrt \pi}e^{-\sigma_{ij}^2r_{ij}^2}\vec\sigma_{i}\cdot\vec\sigma_{j}.
\end{flalign}
The spin-orbital interaction $V_{so}(r_{ij})$ takes the form: 
\begin{flalign}\label{Hsd}
V_{so} &= \frac{1}{r_{ij}} \frac{dV_{V}}{d(r_{ij})} \frac{1}{4} \Bigg[ \left( \frac{1}{m_i^2} + \frac{1}{m_j^2} + \frac{4}{m_i m_j} \right) \vec{L_{ij}} \cdot \vec{S_{ij}} \nonumber \\
& \quad +\left( \frac{1}{m_i^2} - \frac{1}{m_j^2} \right) \vec{L_{ij}} \cdot \left( \vec{s_i} - \vec{s_j} \right) \Bigg] \nonumber \\
& \quad - \frac{1}{r_{ij}} \frac{dV_{S}}{d(r_{ij})} \left( \frac{\vec{L_{ij}} \cdot  \vec{s_i}}{2m_i^2} + \frac{\vec{L_{ij}} \cdot  \vec{s_i}}{m_j^2} \right) \nonumber \\
&=\left( \frac1{m_i^2}+\frac1{m_j^2}+\frac4{m_im_j}\right)(\frac{-B_{ij}\sigma_{ij}}{2\sqrt \pi})\frac{e^{-\sigma_{ij}^2r_{ij}^2}}{r_{ij}^2}\vec{L_{ij}}\cdot\vec S_{ij} \nonumber \\
           &+\left( \frac1{m_i^2}+\frac1{m_j^2}+\frac4{m_im_j}\right)(\frac{-B_{ij}}4)\frac{Erf[{\sigma_{ij}r_{ij}}]}{r_{ij}^3}\vec{L_{ij}}\cdot\vec S_{ij} \nonumber \\
           &+\left(\frac{-A_{ij}}2\right)\frac1{r_{ij}}\left(\frac{\vec L_{ij} \cdot \vec s_i}{m_i^2}+\frac{\vec L_{ij}\cdot \vec s_j}{m_j^2}\right),
\end{flalign}
where $\vec\sigma_i$ are the quark spin operator in SU(2). Note that we have employed $V_V(r)=-B\,Erf[\sigma r]/r$ and $V_S(r)=Ar$, taken from Ref.~\cite{Schoberl:1986bv}. $m_i$ and $m_j$ are the constituent quark masses of the ith and jth quarks.
$\vec{s_i}$ represents the spin operator of the ith quark. $\vec{S_{ij}}=\vec{s_i}+\vec{s_j}$ is the spin operator for (ij)th quark pair. $\vec{L_{ij}}$ is relative orbital operator, taking the form,
\begin{eqnarray}
\vec{L_{ij}}=\vec{r_{ij}}\times\vec{p_{ij}}=\vec{r_{ij}}\times\frac{m_i\vec{p_i}-m_j\vec{p_j}}{m_i+m_j}.
\end{eqnarray}
In line with the previous works, $A_{ij}$, $B_{ij}$, and $\sigma_{ij}$ are proposed to be mass-dependent coupling parameters,  taking the form 
\begin{eqnarray}
A_{ij}= a+bm_{ij},\;\;B_{ij}=B_0 \sqrt{\frac{1}{m_{ij}}},\;\; \sigma_{ij} =\sigma_0{m_{ij}}.
\end{eqnarray}
with $m_{ij}$ being the reduced mass of $i$th and $j$th quarks, defined as $\;m_{ij}=\frac{2 m_i m_j}{m_i+m_j}$. $a$, $b$, $B_0$, and $\sigma_{0}$ are constants. For a more detailed discussion, one may refer to Ref.~\cite{Zhao:2020jvl}. The hyperfine coefficient $\sigma_{ij}$ is also proposed to be mass-dependent~\cite{Schoberl:1986bv}.

The central potential $V_0$ and the spin-spin interaction $V_{ss}$ are treated as the leading effects, while the remaining spin-orbital interaction $V_{so}$ is considered a perturbation that shifts the mass spectrum.
The Schr\"{o}dinger equation is solved for mesons and tetraquarks within the Hamiltonian $H_0$, yielding the eigenvalue $E_0$ and eigenstates $\psi_0$. The mass spectrum is then calculated by adding the diagonalized $H_{so}$ matrix to the basis of the eigenstates obtained in the previous step, which takes the form,
\begin{eqnarray}
E=E_0+\langle \psi_0 | H_{so} | \psi_0 \rangle.
\end{eqnarray}

By importing the coupling constants $a$, $b$, and $B_0$, as well as the constituent quark masses $m_u$, $m_c$, and $m_b$ from previous work~\cite{Zhao:2020jvl}, the mass spectra of S- and P-wave light, charmed, bottom, charmonium, and bottomonium conventional mesons are calculated in the Hamiltonian in Eq. (\ref{eqn::ham}). A comparison of the theoretical results, presented in Table~\ref{tab:groundM}, with experimental data from the PDG~\cite{PDG}, yields a model coupling parameter $\sigma_0 = 0.7$.

{The present calculation is performed in a compact tetraquark CQM basis and does not include an explicit coupling to open two-meson continuum channels. Coupled-channel effects, which may induce additional mass shifts, especially for states close to open-flavor thresholds, will be considered in future extensions of this framework.}

\begin{table}[tbp]
\caption{S- and P-wave meson states applied to fit the model parameters. The last column shows the deviation between the experimental and theoretical mean values, $D=100\cdot (M^{exp}-M^{cal})/M^{exp}$. $M^{exp}$ taken from PDG~\cite{PDG}.}
\label{tab:groundM}
\begin{ruledtabular}
\begin{tabular}{lccccc}

Meson          & $J^{P(C)}$ & nL & $M^{exp}  {\rm (MeV)}$& $M^{cal} {\rm (MeV)}$  & D ($\%$)
\\
 \hline
 $\eta_b$           & $0^{-+}$&  1S   & 9399  & 9394 &  0.1
\\
                          &              &   2S  & 9999  & 10024 & -0.3
 \\
 $\Upsilon$        & $1^{--}$&   1S  & 9460  & 9467   & -0.1
\\
                          &             &   2S & 10023 & 10054 & -0.3
\\
$h_b$                & $1^{+-}$& 1P  & 9899 & 9922 & -0.2
\\
                          &               & 2P  & 10260 & 10320 & -0.6
\\
$\chi_{b0}$        &$0^{++}$&1P   & 9859 & 9896 & -0.4
\\ 
                          &               & 2P  & 10232 & 10301 & -0.7
\\
 $\chi_{b1}$       & $1^{++}$& 1P  & 9893 & 9911 &  -0.2
\\
                          &               & 2P  & 10255 & 10312 & -0.6
\\
 $\chi_{b2}$       & $2^{++}$& 1P  & 9912 & 9935 &  -0.2
\\
                          &               & 2P  & 10269 & 10332 &  -0.6
\\
\hline
$\eta_c$      & $0^{-+}$&  1S & 2984 & 2962 &  0.7
\\
                    &              & 2S  & 3638 & 3614 &  0.7
\\
$\psi$          & $1^{--}$ & 1S & 3097  & 3090 & 0.2
\\
                    &              &  2S & 3686 & 3655 &  0.8
\\
                    &              &  3S & 4040 & 4027 &  0.3
\\
$h_c$          & $1^{+-}$& 1P  & 3525 & 3516 & 0.2
\\
$\chi_{c0}$  &$0^{++}$&1P   & 3415 &  3464 & -1.4
\\
                   &               & 2P  & 3860 & 3876 & -0.4
\\
 $\chi_{c1}$& $1^{++}$& 1P  & 3510 & 3500 &  0.3
\\
 $\chi_{c2}$& $2^{++}$& 1P  & 3556 & 3553 &  0.1
\\
                   &               & 2P  & 3930 & 3932 & -0.1
\\
\hline
 $B^0$       & $0^{-}$ & 1S & 5279 & 5289 & -0.2
\\
 $B^{*}$   & $1^{-}$  & 1S & 5325 & 5337 &  -0.2
\\
$B_1$    & $1^{+}$  & 1P & 5721 & 5800 & -1.4
\\
$B_2^{*}$& $2^{+}$  & 1P & 5747 & 5814 &  -1.2
\\
\hline
 $D^0$       & $0^{-}$ & 1S & 1865 & 1906 & -2.2
\\
                  &              & 2S & 2549 & 2590 &  -1.6
\\
 $D^{*0}$   & $1^{-}$  & 1S & 2007 & 2021 &  -0.7
\\
                  &               & 2S & 2627 & 2622 &  0.2
\\
$D_1^0$    & $1^{+}$  & 1P & 2420 & 2465 & -1.9
\\
$D_0^{*0}$& $0^{+}$  & 1P & 2343 & 2438 & -4.1
\\
$D_1^{0}$& $1^{+}$   & 1P & 2412 & 2465  & -2.2
\\
$D_2^{*0}$& $2^{+}$  & 1P & 2460 & 2508 &  -2.0
\\
\hline
 $\rho$      & $1^{--}$ & 1S  & 770   & 766   &  0.5
 \\
                 &              &  2S  &1450 & 1395 &  3.8
 \\
 $h_1$(1170)     & $1^{+-}$& 1P  & 1170 & \multirow{2}{*}{1202} & -2.7
 \\
 $b_1$(1235)      & $1^{+-}$& 1P  & 1235 &          & 2.3
 \\
 $f_0$(1370)& $0^{++}$& 1P  & 1200-1500 & {1191} &  ...
 \\
 $f_1$(1285)& $1^{++}$& 1P  & 1280 & \multirow{2}{*}{1223} & 4.4
 \\
 $a_1$(1260)      & $1^{++}$& 1P  & 1230 &          & 0.6
 \\
 $f_2$(1270)       & $2^{++}$& 1P  & 1275 & \multirow{2}{*}{1277} & -0.2
 \\
 $a_2$(1320)      & $2^{++}$& 1P  & 1318 &          & 3.1
 \\
\end{tabular}
\end{ruledtabular}
\end{table}

\subsection{Two-body strong decay}

\begin{figure}[tb]
\centering
\includegraphics[width=0.4\textwidth]{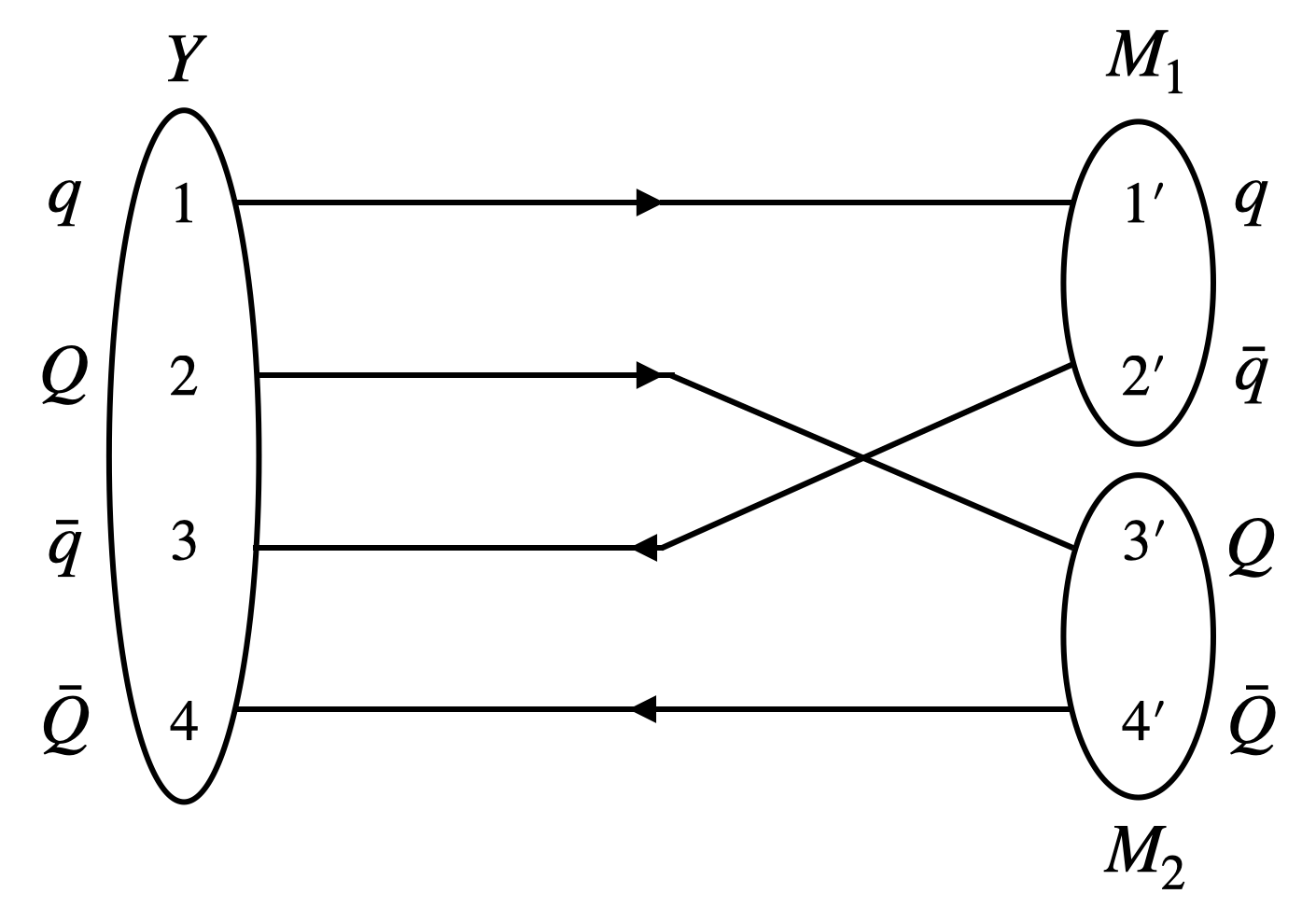}
\caption{Quark rearranged diagram for Y tetraquark in the hidden-charm decay modes.}
\label{fig:QRD}
\end{figure}

The two-body strong decays of the Y tetraquark states are studied within the rearrangement mechanism, as shown in Figure~\ref{fig:QRD}. 
\begin{eqnarray}
T_{cs}&=&\langle \psi_f^{cs} | \psi_i^{cs} \rangle  \nonumber \\
&=&\langle \psi_{M_1M_2}^{c} | \psi_Y^{c} \rangle \langle \psi_{M_1M_2}^{s} | \psi_Y^{s} \rangle
\end{eqnarray}
where $\psi_Y^{c}$ and $\psi_Y^{s}$ are color and spin wave functions of the initial Y tetraquark states in the mass spectrum. $\psi^c_{M_1M_2}$ and $\psi^s_{M_1M_2}$ are the color and spin wave functions of the final states of all possible two-body strong decay channels: $\omega \chi_{c0}$, $\omega \chi_{c1}$, $\omega \chi_{c2}$, $\eta J/\psi$, $\omega \chi_{b0}$, $\omega \chi_{b1}$, and $\omega \chi_{b2}$. 

In the work, the ratio of $|T_{cs}|^2$ is approximated as the branching ratio, assuming that the spatial contributions from different decay channels are similar.

\section{\label{sec:RND}Results and Discussion}

\subsection{Results}

\begin{table}[tbp]
\caption{Mass spectrum and decay branching ratios for the $S=0$ $Y_c$ tetraquark states. Threshold masses are given in square brackets for each decay channel, and the ratios are normalized to the $E2\to\omega \chi_{c0}$ channel (set to 1). All masses are in MeV.}
\label{tab:s0tetra}
\begin{ruledtabular}
\begin{tabular}{lcccccc}
En   &  Mass  & $\omega \chi_{c0}$[4196] & $\omega \chi_{c1}$[4292] & $\omega \chi_{c2}$[4338] 
\\
\hline
E1   & 4141    &  ...     &  ...       &   ...      
\\
E2   & 4220    &  1  &  ...   &  ...   
\\
E3   & 4301    &  15     &  45     &  ...      
\\
E4   & 4314    &  1.8    &  5   &  ...    
\\
E5   & 4390    &  1.7    &   5  &  8.5   
\\
E6   & 4409    &  0.01      &  0.03     &   0.05       
\\
E7   & 4482    &  4.8    &  14     &   24    
\\
E8   & 4499    &  0.3    & 0.7      &   1.3     
\\
E9   & 4518    &  6    & 18    &   30  
\\
E10 & 4532    &  1.4    & 4      &  7    
\\
E11 & 4567    &  6    & 17    &  28    
\\
E12 & 4576    &  0.3     & 1      & 1.6      
\\
E13 & 4588    &  11  & 33   & 55    
\\
E14 & 4597    &  3   & 10    &  17   
\\
E15 & 4626    &  56     &  169     &  282          
\\
\end{tabular}
\end{ruledtabular}
\end{table}

\begin{table}[tbp]
\caption{{Mass spectrum and decay branching ratios for the $S=0$ $Y_b$ tetraquark states. Threshold masses are given in square brackets for each decay channel, and the ratios are normalized to the $E4\to\omega \chi_{b0}$ channel (set to 1). All masses are in MeV.}}
\label{tab:bbs0tetra}
\begin{ruledtabular}
\begin{tabular}{lcccccc}
En   &  Mass  & $\omega \chi_{b0}$[10641] & $\omega \chi_{b1}$[10674] & $\omega \chi_{b2}$[10694]
\\
\hline  
E1   & 10504    &  ...     &  ...      &   ...      
\\
E2   & 10548    &  ...     &  ...      &  ...   
\\
E3   & 10565    &  ...     &  ...      &  ...     
\\
E4   & 10679    &  1      &  3    &  ...   
\\
E5   & 10754    &  1.4    &   4      &   7   
\\
E6   & 10774    &  0.1      &   0.3      &   0.5    
\\
E7   & 10791    &  0.2    &   0.7     &   1.2    
\\
E8   & 10810    &  0.2   &   0.5      &   0.8    
\\
\end{tabular}
\end{ruledtabular}
\end{table}

\begin{table}[tbp]
\caption{{Mass spectrum of mixed $S=2$ $Y_c$ and $Y_b$ tetraquark states. Mass values are in MeV.}}
\label{tab:s2tetra}
\begin{ruledtabular}
\begin{tabular}{lcccccccc}
En & E1   &  E2  & E3 & E4 & E5 & E6
\\
\hline 
$Y_c$ & 4285 & 4351 & 4464 & 4571 & 4579 & 4618
\\
$Y_b$ & 10574 & 10657 & 10797 & 10830 & 10855 & 10940
\\
\end{tabular}
\end{ruledtabular}
\end{table}

The predetermined and imported parameters are applied to predict the masses of Y tetraquark states in the Hamiltonian in Eq.~(\ref{eqn::ham}) including the color-dependent central potential $V_{cen}$, which may mix different color-spin configurations.

Because of the cross terms from the color operator $(\vec\lambda_{i}\cdot\vec\lambda_{j})$ and the spin operator $(\vec\sigma_{i}\cdot\vec\sigma_{j})$, listed in Appendix~\ref{sec:AP1}, eigenstates of the Hamiltonian for S=0 are linear combinations of $\psi^c_{\bar 3\otimes3}\psi^{S=0}_{(0\otimes0)}$, $\psi^c_{\bar 3\otimes3}\psi^{S=0}_{(1\otimes1)}$, $\psi^c_{6\otimes\bar 6}\psi^{S=0}_{(0\otimes0)}$, and $\psi^c_{6\otimes\bar 6}\psi^{S=0}_{(1\otimes1)}$. Eigenstates for $S=2$ are linear combinations of $\psi^c_{\bar 3\otimes3}\psi^{S=2}_{(1\otimes1)}$ and $\psi^c_{6\otimes\bar 6}\psi^{S=2}_{(1\otimes1)}$.
There is no configuration mixing between $S=0$ and $S=2$ states, as no cross terms are found. En, where $n = 1, 2, 3, \dots$, represents the $n$th eigenstate of the Hamiltonian in Eq.~(\ref{eqn::ham}) in the notation used later in this work, incorporating color-spin configuration mixing.

The theoretical masses and branching ratios for the E1--E16 $S=0$ charmonium-like $Y_c$ tetraquarks with various mixed configurations are listed in Table~\ref{tab:s0tetra}. 
Similarly, the masses and branching ratios of the E1 to E8 mixed $S=0$ bottomonium-like $Y_b$ tetraquarks are listed in Table~\ref{tab:bbs0tetra}.
Likewise, the masses of the E1 to E6 mixed $S=2$ charmonium-like $Y_c$ and bottomonium-like $Y_b$ tetraquarks are provided in Table~\ref{tab:s2tetra}.

{For $S=0$ $Y_c$ tetraquarks, the spin-color factor for the two-body strong decay into $\eta J/\psi$ is close to zero.}
For $S=2$ $Y_c$ and $Y_b$ tetraquarks, the spin-color factors in $\omega \chi_{c0}$, $\omega \chi_{c1}$, $\omega \chi_{c2}$, $\omega \chi_{b0}$, $\omega \chi_{b1}$, $\omega \chi_{b2}$ decay channels are all equal to zero. 

\begin{figure}[tb]
\centering
\includegraphics[width=0.47\textwidth]{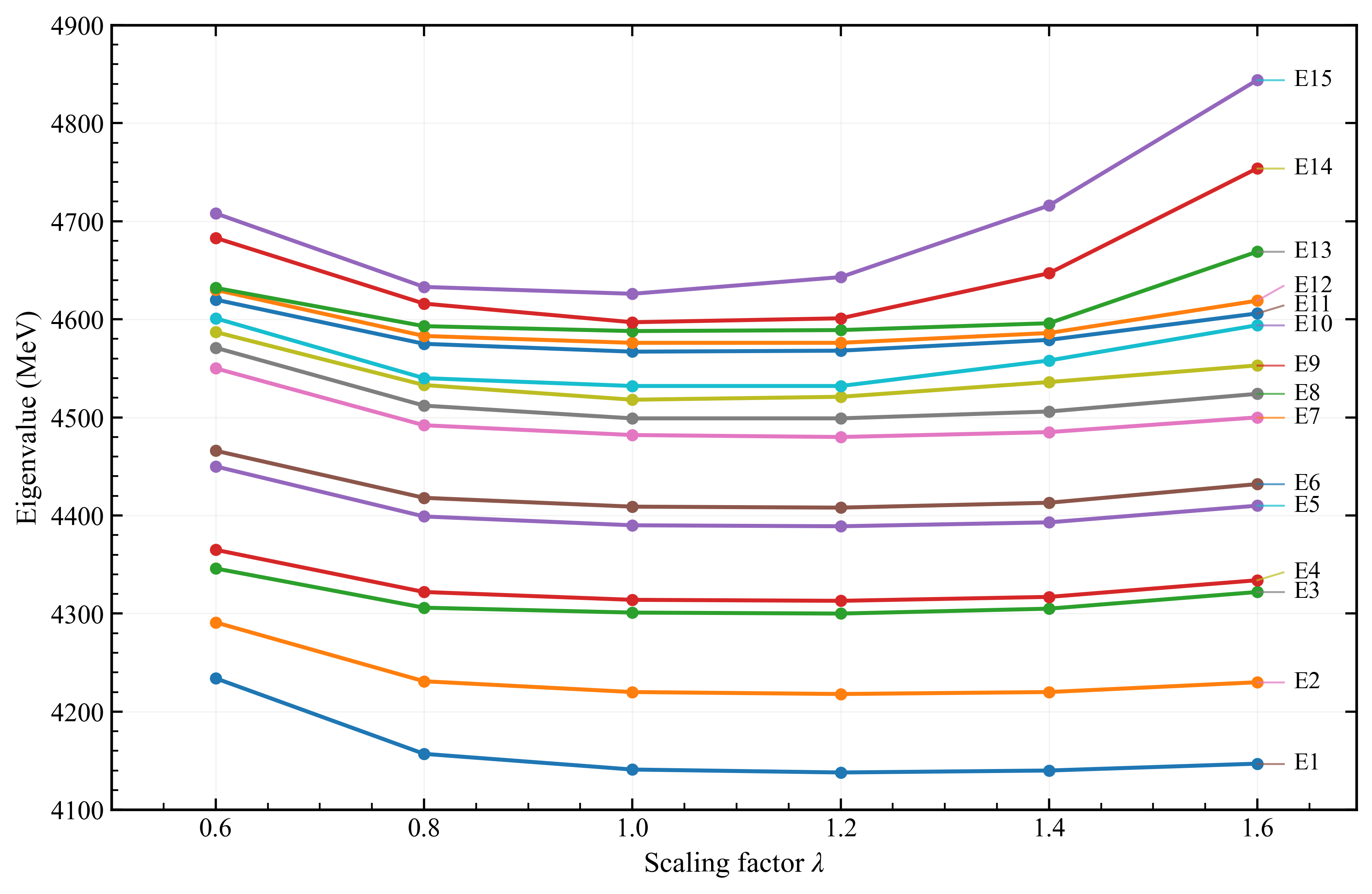}
\caption{{Real-scaling trajectories of the mixed $S=0$ $Y_c$ tetraquark levels. The eigenvalues $E_n(\lambda)$ (E1--E15) are obtained by rescaling the length parameters in all Jacobi coordinates as $b\to\lambda b$ while keeping the Hamiltonian fixed.}}
\label{fig:RSG2M}
\end{figure}

{Since the diquark-type color couplings $\bar 3_c\!-\!3_c$ and $6_c\!-\!\bar 6_c$ are linearly related to the meson--meson couplings $1_c\!-\!1_c$ and $8_c\!-\!8_c$, an eigenstate may occasionally acquire a noticeable $1_c\!-\!1_c$ admixture. Such a component can resemble a weakly interacting two-hadron configuration and may therefore manifest a stronger basis sensitivity. We have applied the real-scaling method~\cite{Richard:2024cfg} to verify the stability of each eigenstate against changes of the basis length parameter and, in particular, to flag possible continuum-like behavior associated with a sizeable $1_c\!-\!1_c$ (two color-singlet cluster) component. The length parameters $b$ in the basis are rescaled by a common factor, $b\to\lambda b$, while the basis structure and Hamiltonian are kept fixed~\cite{Hiyama:2018ukv,Hu:2022pae,Tan:2023azs}.}

{We present, as an example, in Fig.~\ref{fig:RSG2M} the eigenvalue trajectories of the $S=0$ $Y_c$ levels, E1--E15, over a representative range of $\lambda$. The low-lying levels E1--E12 display clear flat valleys, with variations only at the level of a few tens of MeV. The highest excitations (E13--E15) shift more visibly, indicating a stronger sensitivity to the basis truncation.}

Experimental data on $Y_c$ and $Y_b$ state candidates from different decay processes are listed in Table~\ref{tab:data} and will be reviewed and discussed separately below.
The assignments of $S=0$ and $S=2$ $Y_c$ tetraquarks, along with their theoretical masses and two-body decay branching ratios to $\omega \chi_{c0}$ are presented in Table~\ref{tab:ass}.
Similarly, the assignments of $S=0$ $Y_b$ tetraquarks, along with their theoretical masses and two-body decay branching ratios to $\omega \chi_{b0}$, $\omega \chi_{b1}$, and $\omega \chi_{b2}$, are provided in Table~\ref{tab:assbb}.

\begin{table*}[htbp]
\caption{\label{tab:data} Masses and widths of $Y_c$ and $Y_b$ state candidates from the cited sources, with units in MeV. PDG represents the names of established states in the PDG, while Assignment refers to the names used in this work. The symbol $\psi$ in the third column represents both $J/\psi$ and $\psi(2S)$.}
\begin{ruledtabular}
\begin{tabular}{ccllllllc}
PDG   &   Assignment     &  $\pi^+\pi^-\psi$                   &  $K^+K^-J/\psi$                          & $\omega\chi_{cJ}$  &  $\pi^+\pi^-h_c$  &  $\eta J/\psi$ & $\pi^+\pi^-\Upsilon(nS)$ &  $\omega\chi_{bJ}$
\\
\hline
{$\psi(4230)$} &{Y(4230)} & $4220$, $46$ \cite{BESIII:2020oph}  & $4225$, $73$ \cite{BESIII:2022joj} & $4218$, $28$  \cite{BESIII:2019gjc} & $4218$, $66.0$ \cite{BESIII:2016adj} & $4219$, $82.7$ \cite{BESIII:2020bgb}
\\               
                    && $4221$, $42$ \cite{BESIII:2022qal}
\\
                    && $4234$, $18$ \cite{BESIII:2021njb}
\\
\hline
{$\psi(4360)$} &{Y(4300)} & $4298$, $127$ \cite{BESIII:2022qal} 
\\ 
\cline{2-9}
                    &{Y(4340)} & $4340$, $94$ \cite{BaBar:2012hpr}
\\
                    && $4347$, $103$ \cite{Belle:2014wyt}
\\
 \cline{2-9}

  & {Y(4390)}  & $4390$, $143$ \cite{BESIII:2021njb} &  &  &  $4392$, $140$ \cite{BESIII:2016adj} & $4382$, $136$ \cite{BESIII:2020bgb} & 
\\
\hline
... & Y(4484) & & $4484$, $111$ \cite{BESIII:2022joj} & & & 
\\ 
\hline
... &Y(4544) & & & $4544$, $116$ \cite{BESIII:2024jzg}
\\                
\hline
{$\psi(4660)$}&{Y(4660)} & $4651$, $155$ \cite{BESIII:2021njb} & $4708$, $126$ \cite{BESIII:2023wqy} &  & &  & &  
\\
                    &&  $4652$, $68$ \cite{Belle:2014wyt}
\\
                    &&  $4669$, $104$ \cite{BaBar:2012hpr}
\\
\hline
$\Upsilon(10753)$ &Y(10753) & & & & & &  $10753$, $36$~\cite{Belle:2019cbt} & {Yes}~\cite{Belle-II:2022xdi}

\end{tabular}
\end{ruledtabular}
\end{table*}

\begin{table}[hptb]
\caption{\label{tab:ass} Presenting theoretical mass and branching ratio predictions, along with assignments for $S=0$ and $S=2$ $Y_c$ tetraquark states. The mass results are given in units of MeV. The Y(4390) and Y'(4390) denote the states observed in the $\pi^+\pi^-h_c$ and $\pi^+\pi^-\psi$ processes, respectively.}
\small  
\begin{ruledtabular}
\begin{tabular}{l@{\hspace{1pt}}c@{\hspace{1pt}}c@{\hspace{1pt}}c@{\hspace{1pt}}c@{\hspace{1pt}}c@{\hspace{1pt}}c@{\hspace{1pt}}c@{\hspace{1pt}}c}
Mass                       & 4141      & 4220      & 4285      & 4301       & 4314   & 4351      & 4390 \\
\hline
$\omega \chi_{c0}$ & ...           & 1            & ...           & 15           & 1.8      & ...           & 1.7  \\
Data                        & X(4160) & Y(4230)  & Y(4300) & ...            & ...        & Y(4340) & Y(4390) \\
\hline
\hline
Mass                        & 4409     & 4464   & 4482        & 4499       & 4518 & 4532 & 4567      \\
\hline
$\omega \chi_{c0}$ & 0.01       & ...        & 4.8           & 0.3          & 6       & 1.4    & 6         \\
Data                        & Y'(4390) & ...       &  ...             & Y(4484)  &  ...     & ...      & Y(4544) \\
\hline
\hline
Mass                       & 4571      & 4576   & 4579      & 4588 &  4597  & 4618        & 4626        \\ 
\hline
$\omega \chi_{c0}$ & ...           & 0.3      & ...           & 11     & 3         & ...             & 56             \\ 
Data                        & ...           & ...        &       ...     &      ... & ...        & Y(4660)   & ...              \\ 

\end{tabular}
\end{ruledtabular}
\end{table}

\subsection{Y(4230)}

Over the past two decades, the Y(4230) has been established as an exotic state observed and confirmed by several experimental collaborations, and is considered part of the broader family of charmonium-like states~\cite{PDG}. 
{The Y(4260), the original name of Y(4230), was observed by BaBar in the $\pi^+\pi^-J/\psi$ invariant-mass spectrum~\cite{BaBar:2005hhc}, and was subsequently confirmed by CLEO Collaboration~\cite{CLEO:2006tct} and Belle Collaboration~\cite{Belle:2007dxy} in the same process. 
With improved statistical precision, BESIII identified an asymmetry in the cross section, which caused the peak position to move to a lower mass~\cite{BESIII:2016bnd}. The Y(4260) was therefore renamed Y(4230). }

Unlike conventional quarkonium, which consists of a quark-antiquark pair, the Y(4230) is classified as an exotic hadron due to its unconventional properties, including anomalous decay channels and unexpected production mechanisms. In the previous work, the Y(4230) could not be accommodated in a conventional charmonium picture including $S-D$ mixing~\cite{Zhao:2023hxc}.

The exact internal structure of the Y(4230) remains a subject of ongoing research, with theoretical models proposing interpretations such as tetraquark states~\cite{Maiani:2005pe,Ali:2017wsf,Wang:2018ejf}, molecular bound states of charmed mesons~\cite{Chiu:2005ey,Qin:2016spb}, or hybrid mesons~\cite{Oncala:2017hop,Berwein:2015vca}. 

The Y(4230) has been observed in several channels, but primarily via electron-positron collisions, where the processes include $e^+e^-\to\pi^+\pi^-\psi$\cite{BESIII:2020oph,BESIII:2022qal,BESIII:2021njb}, $\eta J/\psi$\cite{BESIII:2020bgb}, $\omega\chi_{c0}$\cite{BESIII:2014rja}, $\pi^+\pi^-h_c$\cite{BESIII:2016adj}, and $K^+K^-J/\psi$\cite{BESIII:2022joj}, which are listed in Table~\ref{tab:data}. Observations in different channels provide clues for identifying whether certain resonances correspond to distinct states.

{In the first observation of Y(4230) in the invariant mass of $\omega\chi_{c0}$ by BESIII in 2014~\cite{BESIII:2014rja}, the resonance parameters were found to be inconsistent with the line shape of the Y(4260) observed in $\pi^+\pi^-J/\psi$\cite{BaBar:2005hhc} which was the first observed Y state by BaBar in 2005. 
However, the Y(4260) structure in the $\pi^+\pi^-J/\psi$ channel was later superseded in PDG by an updated BaBar analysis~\cite{BaBar:2012vyb}, in which the presence of two interfering resonances was not excluded. 
Similarly, the Y(4230) observed in the $\omega\chi_{c0}$ channel~\cite{BESIII:2014rja} was superseded by an updated BESIII analysis in 2019~\cite{BESIII:2019gjc}. In this updated BESIII study, no solid conclusion could be drawn regarding whether the Y(4230) observed in the $\pi^+\pi^-J/\psi$ and $\omega\chi_{c0}$ channels corresponds to the same state.}

The Y(4230) was also observed in the $K^+K^-J/\psi$ process~\cite{BESIII:2022joj}, and the resonance parameters are consistent with those of the Y(4230) observed in the $\pi^+\pi^-J/\psi$ channel, which suggests that the signals seen in $K^+K^-J/\psi$ and $\pi^+\pi^-J/\psi$ may correspond to the same state.
Later, the Y(4230) was observed in the process $\pi^+\pi^-h_c$~\cite{BESIII:2016adj}, whose resonance parameters are consistent with those of the resonance observed in $\omega\chi_{c0}$~\cite{BESIII:2014rja}, which supports the possibility that the structures seen in $\omega\chi_{c0}$ and $\pi^+\pi^-h_c$ correspond to the same state.

The number of states in this mass region remains to be further determined experimentally~\cite{PDG}. This study predicts only one state in the mass region. {As shown in Table~\ref{tab:ass}, we assign the Y(4230), observed in the $\pi^+\pi^-J/\psi$, $K^+K^-J/\psi$, $\omega\chi_{c0}$, and $\pi^+\pi^-h_c$ channels, to the E2 tetraquark state, with a corresponding theoretical mass of 4220 MeV.}

\subsection{Y(4360)}

The Y(4360) was first observed in the process of $e^+e^- \to \gamma \pi^+ \pi^- \psi(2S)$~\cite{Belle:2007umv}. With a mass of approximately 4.36 GeV, Y(4360) resides in the energy region above the open-charm threshold, similar to other Y-states. The Y(4360) is of particular interest due to its unusual decay patterns, including prominent decays into $\psi(2S)$ rather than into lower charmonium states like $J/\psi$, which contrasts with expectations from conventional quarkonium models. This anomalous behavior suggests that Y(4360) could be an exotic hadron.

As a PDG established state, in PDG data listed in Table~\ref{tab:data}, the reported masses of $\psi(4360)$ range from 4.3 to 4.4 GeV, which may indicate three distinct states within this broad mass range. 
{The first, second, and third states are around 4.30 GeV, 4.34 GeV, and 4.39 GeV, and are denoted here as Y(4300), Y(4340), and Y(4390), respectively. }

{The Y(4300) and Y(4340) observed in $\pi^+\pi^-J/\psi$ and $\pi^+\pi^-\psi(2S)$ processes are assigned as $S=2$ tetraquark states, with a corresponding theoretical mass of 4285 MeV and 4351 MeV, respectively.}

{The Y(4390) was observed in $\pi^+\pi^-J/\psi$~\cite{BESIII:2021njb} and $\pi^+\pi^-h_c$~\cite{BESIII:2016adj} channels. In the $\pi^+\pi^-h_c$ process~\cite{BESIII:2016adj}, the Y(4230) was also observed. The resonance parameters of the Y(4230) in $\pi^+\pi^-h_c$ process were consistent with those of the Y(4230) observed in the $\omega\chi_{c0}$ channel~\cite{BESIII:2014rja}, which may suggest that the Y(4390) could also be observed in the $\omega\chi_{c0}$ channel.}

{The experimentally observed Y(4390) in the $\pi^+\pi^-h_c$ channel is assigned to the state with a theoretical mass of 4390 MeV in Table~\ref{tab:ass}, considering the large decay ratio to $\omega\chi_{c0}$.} The Y(4390) could potentially be observed in the $\omega\chi_{c0}$ channel, which may motivate future experimental searches.

{In Ref.~\cite{BESIII:2016adj}, the line shapes of Y(4390) observed in $\pi^+\pi^-h_c$ process~\cite{BESIII:2016adj} are inconsistent with those in $\pi^+\pi^-J/\psi$ and $\pi^+\pi^-\psi(2S)$ by BaBar~\cite{BaBar:2005hhc,BaBar:2006ait,BaBar:2012hpr,BaBar:2012vyb} and by Belle~\cite{Belle:2007dxy,Belle:2007umv,Belle:2014wyt,Belle:2013yex}. One may support the existence of two Y states around 4390 MeV, which are tentatively named Y(4390) for the one from $\pi^+\pi^-h_c$, and Y'(4390) for the one from $\pi^+\pi^-J/\psi$.}

{In Table~\ref{tab:ass}, the experimentally observed Y'(4390) in the $\pi^+\pi^-\psi$ channel is assigned to the state with a theoretical mass of 4409 MeV, due to the small decay ratio to $\omega\chi_{c0}$.}

\subsection{Y(4660)}

The Y(4660) was first discovered in the initial state radiation process $ e^+e^- \to \gamma_{\text{ISR}} \pi^+ \pi^- \psi(2S)$~\cite{BaBar:2012hpr}. With a mass of approximately 4.66 GeV, Y(4660) lies well above the open-charm threshold, adding complexity to the spectrum of charmonium-like states. The Y(4660) exhibits intriguing decay patterns, most notably decaying into $\psi(2S)$ rather than $J/\psi$, a feature shared with other exotic states like Y(4360). 
This has led to speculation that the Y(4660) could be a tetraquark~\cite{Lebed:2016yvr,Lu:2017yhl,Wang:2018ejf}, a $S-D$ wave mixture state~\cite{Bhavsar:2018umj}, or a hybrid meson~\cite{Oncala:2017hop,Berwein:2015vca}.

{As shown in Table~\ref{tab:ass}, the experimentally observed Y(4660) has been reported mainly in the $\pi^+\pi^-\psi$ channel and may be naturally grouped with Y(4300), Y(4340), and Y'(4390), which are seen in the same process. The Y(4660) may be assigned to a tetraquark state with a small decay ratio to $\omega\chi_{cJ}$ and a corresponding theoretical mass of 4618 MeV.}

\subsection{Other $Y_c$ states}

The X(4160) was first observed by the Belle Collaboration in double-charmonium production, $e^+e^- \to J/\psi X(4160)$ with $X(4160)\to D^{+}D^{-}$\cite{Belle:2007woe}. Later, a related state was observed by the LHCb Collaboration in the decay $B^+ \to J/\psi\phi K^+$\cite{LHCb:2021uow}, reconstructed from the invariant mass spectrum of $J/\psi\phi$.
The two states observed around 4.15 GeV are grouped together as a single entry in the PDG~\cite{PDG}, according to their similar masses and decay widths. Whether these observations correspond to the same state still requires further experimental confirmation. The nature of the X(4160) therefore remains uncertain and requires further theoretical and experimental investigation.

The production rate of the X(4160) in double-charmonium processes, along with its decay patterns into the $D^{+}D^{-}$ and $J/\psi \phi$ channels, does not align well with expectations for conventional $c\bar{c}$ states~\cite{Chao:2007it,PDG}. However, some studies have still considered interpretations of the X(4160) as a higher charmonium state~\cite{Yang:2009fj}.
The X(4160) is more likely interpreted as an exotic hadron, with several theoretical scenarios proposed. These include a compact tetraquark configuration, either hidden-charm $c\bar{c}q\bar{q}$ or $c\bar{c}s\bar{s}$, supported by QCD sum rules~\cite{Chen:2017dpy}, as well as a meson-meson molecular state, in particular a $D_s^*\bar D_s^*$ bound state~\cite{Wang:2017mrt}.

Despite these scenarios, the quantum numbers of the X(4160) remain unclear. The Belle Collaboration did not measure its quantum numbers~\cite{Belle:2007woe}, while LHCb analyses favor an assignment of $J^{PC}=2^{-+}$~\cite{LHCb:2021uow}. Given its measured mass near 4.16 GeV, the X(4160) may also be interpreted as a tetraquark candidate, based on mass consistency with theoretical predictions for hidden-charm exotics in this region~\cite{Chen:2017dpy}. However, further experimental confirmation is still required. X(4160) is tentatively assigned as a tetraquark state based on mass agreement, with a corresponding theoretical mass of 4141 MeV.

Recently, Y(4484) was observed in the $K^+K^-J/\psi$ process by BESIII for the first time with a statistical significance greater than 8$\sigma$~\cite{BESIII:2022joj}. Together with the Y(4230) observed in the same process, the line shape is consistent with that of the established Y(4230) in the $\pi^+\pi^-J/\psi$ channel. {In Table~\ref{tab:ass}, the Y(4484) is grouped with Y'(4390) from $\pi^+\pi^-J/\psi$ channel and is assigned to be a tetraquark state based on mass agreement, with a corresponding theoretical mass of 4499 MeV.}

The newly observed Y(4544) was seen from $\omega\chi_{c1}$ channel by BESIII~\cite{BESIII:2024jzg}, and the structure was observed for the first time with a significance of 5.8$\sigma$. The Y(4544) has significantly higher mass compared to the Y(4484). One may support the view that Y(4484) and Y(4544) are different states rather than combining them into a single Y(4500). {In the assignment, the Y(4544) is grouped with Y(4230) from $\omega\chi_{c0}$ channel and assigned to be a tetraquark state considering the mass agreement and the large ratio to $\omega\chi_{cJ}$ with a corresponding theoretical mass of 4567 MeV.}

{The theoretical branching ratios for $1^{--}$ Y tetraquark states decaying into $\eta J/\psi$ are close to zero.} Thus, the Y(4230) state and the Y(4360) state observed from $\eta J/\psi$~\cite{BESIII:2020bgb} cannot be accommodated in the current tetraquark picture, and these two states could correspond to other structures, for example, hadronic molecule states~\cite{Chiu:2005ey,Qin:2016spb,Ding:2008gr,Close:2009ag,Close:2010wq}, or hybrid mesons~\cite{HadronSpectrum:2012gic,Luo:2005zg,Zhu:2005hp,Oncala:2017hop,Berwein:2015vca}.

\subsection{Y(10753)}

\begin{table}[hptb]
\caption{\label{tab:assbb} Theoretical mass, branching ratio predictions, and assignments for $S=0$ $Y_b$ tetraquark states. Mass values are in MeV.}
\small  
\begin{ruledtabular}
\begin{tabular}{l@{\hspace{3pt}}c@{\hspace{3pt}}c@{\hspace{3pt}}c@{\hspace{3pt}}c@{\hspace{3pt}}c@{\hspace{3pt}}c@{\hspace{3pt}}c@{\hspace{3pt}}c} 
En               & E1   & E2   & E3   & E4   & E5   & E6   & E7   & E8   \\ 
\hline        
Mass            & 10504 & 10548 & 10565 & 10679 & 10754 & 10774 & 10791 & 10810 \\ 
$\omega \chi_{b0}$ & ...    & ...  & ...   & 1   & 1.4 & 0.1   & 0.2   & 0.2   \\ 
$\omega \chi_{b1}$ & ...    & ...  & ...   & 3   & 4    & 0.3   & 0.7   & 0.5   \\ 
$\omega \chi_{b2}$ & ...    & ...  & ...   & ...  & 7    & 0.5   & 1.2   & 0.8   \\ 
Data     &  &  &  &  & Y(10753)  &  &  &  \\ 
\end{tabular}
\end{ruledtabular}
\end{table}

The Y(10753) bottomonium-like state was first observed by Belle in the $e^+e^- \to \pi^+\pi^- \Upsilon(nS)$ process~\cite{Belle:2019cbt}. {Recently, the Belle II Collaboration reported another observation that strongly supports the existence of the radiative transition process $\Upsilon(10753) \to \omega\chi_{bJ}(1P)$~\cite{Belle-II:2022xdi}. The energy dependence of the Born cross sections for Y(10753) in the $\omega\chi_{bJ}(1P)$ channel~\cite{Belle-II:2022xdi} was found to be consistent with the shape of the Y(10753) state in the $\pi^+\pi^- \Upsilon(nS)$ channel~\cite{Belle:2019cbt}.} Belle II concluded that the internal structure of Y(10753) may differ from that of $\Upsilon(10860)$, where the latter is well understood as a predominantly conventional bottomonium state. 

In previous work, the Y(10753) could not be accommodated within a conventional bottomonium picture, including $S-D$ mixing~\cite{Zhao:2023hxc}. After the experimental discovery, a P-wave hidden-bottom tetraquark interpretation was proposed in the first set of theoretical pictures~\cite{Wang:2019veq}. For a tetraquark mixing interpretation, one may refer to Ref~\cite{Ali:2019okl}. 

As shown in Table~\ref{tab:assbb}, the experimentally observed Y(10753) in $\pi^+\pi^- \Upsilon(nS)$ and $\omega\chi_{bJ}(1P)$ channels is assigned to the E5 tetraquark state, considering the mass agreement and the large ratio to $\omega\chi_{bJ}$. 

\section{Summary}\label{sec:SUM}

The masses of P-wave charmonium-like tetraquark states are calculated using a constituent quark model (CQM) that incorporates a Cornell-like potential and the Breit-Fermi interaction. All model parameters were predetermined by reproducing the mass spectra of low-lying S- and P-wave light, charmed, and charmonium mesons. The theoretical results for P-wave tetraquarks are compared with selected exotic states, known as Y states, and a tentative assignment is proposed.

The PDG-established state around 4.23 GeV may be treated as a single state at this stage. In the assignment, the Y(4230), observed in the $\pi^+\pi^-J/\psi$, $K^+K^-J/\psi$, $\omega\chi_{c0}$ and $\pi^+\pi^-h_c$ channels, is assigned as an $S=0$ $1^{--}$ tetraquark state.

The PDG established state around 4.36 GeV may split into four distinct states. 
In the assignment, the Y(4300) and Y(4340) observed only in the $\pi^+\pi^-J/\psi$ process are assigned as $1^{--}$ tetraquark states, respectively. 
The Y(4390), observed in the $\pi^+\pi^-h_c$ channel, is assigned as a $1^{--}$ tetraquark state with a large decay ratio to $\omega\chi_{cJ}$, while the Y'(4390), observed in the $\pi^+\pi^-J/\psi$ channel, is assigned as a tetraquark state with a small decay ratio to $\omega\chi_{cJ}$.
The Y(4390) could potentially be observed in the $\omega\chi_{c0}$ channel, which may motivate future experimental searches.

The PDG established state Y(4660), observed in the $\pi^+\pi^-J/\psi$ channel and grouped with Y'(4390) from the same process, is assigned as a $1^{--}$ tetraquark state. 

The Y(4484), observed in the $K^+K^-J/\psi$ channel by BESIII, and the Y(4544), newly observed in the $\omega\chi_{c1}$ channel by BESIII,
are tentatively assigned as $1^{--}$ tetraquark states because their masses agree with our predictions. 

The Y(10753), observed in $\pi^+\pi^- \Upsilon(nS)$ and $\omega\chi_{bJ}(1P)$ channels by Belle, is assigned as the E5 $S=0$ $1^{--}$ bottomonium-like tetraquark state. 

\begin{acknowledgments}
This research has received funding support from the NSRF via the Program Management Unit for Human Resources \& Institutional Development, Research and Innovation [grant number~B50G670107].
A.~L. and Y.~Y. acknowledge the support of (i) Suranaree University of Technology, (ii) Thailand Science Research and Innovation (TSRI), and (iii) NSRF, Project No.~195242.
Z.~Z. is additionally supported by SUT (Full-Time 66/15/2024).
\end{acknowledgments}

\appendix

\section{Tetraquark spatial wave function}\label{sec:AP2}

\indent The total spatial wave function of the tetraquark, coupling the $\chi_1$, $\chi_2$ and $\chi_3$ HO wave functions, takes the general form,
\begin{eqnarray}\label{eqn::spatial}
\psi_{NL} 
&=\sum_{{n_{\chi_i},l_{\chi_i}}}
 A(n_{\chi_1},n_{\chi_2},n_{\chi_3},l_{\chi_1},l_{\chi_2},l_{\chi_3}) \nonumber \\
&\times \psi_{n_{\chi_1}l_{\chi_1}}(\vec \chi_1\,) \otimes\psi_{n_{\chi_2}l_{\chi_2}}(\vec \chi_2\,)\otimes\psi_{n_{\chi_3}l_{\chi_3}}(\vec \chi_3)
\end{eqnarray}
where $\psi_{n_{\chi_i}l_{\chi_i}}$ are HO wave functions. The sum ${n_{\chi_i},l_{\chi_i}}$ is over $n_{\chi_1},n_{\chi_2},n_{\chi_3}, l_{\chi_1},l_{\chi_2},l_{\chi_3}$.
$N$ and $L$ are respectively the total principal quantum number and total angular momentum, 
and $L=l_{\chi_1}+l_{\chi_2}+l_{\chi_3}=1$ while $l_{\chi_1},\,l_{\chi_2},\,l_{\chi_3}\leq1$ for P-wave tetraquark and $N= (2n_{\chi_1}+ l_{\chi_1})+(2n_{\chi_2}+ l_{\chi_2})+(2n_{\chi_3}+l_{\chi_3})$. 

The complete bases of the tetraquarks are listed in Table \ref{three1} up to $N=3$ for reference.

\begin{table}[htbp]
\caption{\label{three1} The complete bases of tetraquark with quantum number, $N=2n+L$ and $L=1$.}
\begin{ruledtabular}
\begin{tabular}{@{}ll}
$\psi_{11}$ & $\psi_{0,0}(\vec\chi_1\,)\psi_{0,0}(\vec\chi_2\,)\psi_{0,1}(\vec\chi_3\,)$, $\psi_{0,0}(\vec\chi_1\,)\psi_{0,1}(\vec\chi_2\,)\psi_{0,0}(\vec\chi_3\,)$, \\
\phantom{} & $\psi_{0,1}(\vec\chi_1\,)\psi_{0,0}(\vec\chi_2\,)\psi_{0,0}(\vec\chi_3\,)$ \\ 
$\psi_{31}$ & $\psi_{1,0}(\vec\chi_1\,)\psi_{0,0}(\vec\chi_2\,)\psi_{0,1}(\vec\chi_3\,)$, $\psi_{1,0}(\vec\chi_1\,)\psi_{0,1}(\vec\chi_2\,)\psi_{0,0}(\vec\chi_3\,)$, \\ 
\phantom{} & $\psi_{1,1}(\vec\chi_1\,)\psi_{0,0}(\vec\chi_2\,)\psi_{0,0}(\vec\chi_3\,)$, $\psi_{0,0}(\vec\chi_1\,)\psi_{1,0}(\vec\chi_2\,)\psi_{0,1}(\vec\chi_3\,)$, \\
\phantom{} & $\psi_{0,0}(\vec\chi_1\,)\psi_{1,1}(\vec\chi_2\,)\psi_{0,0}(\vec\chi_3\,)$, $\psi_{0,1}(\vec\chi_1\,)\psi_{1,0}(\vec\chi_2\,)\psi_{0,0}(\vec\chi_3\,)$, \\
\phantom{} & $\psi_{0,0}(\vec\chi_1\,)\psi_{0,0}(\vec\chi_2\,)\psi_{1,1}(\vec\chi_3\,)$, $\psi_{0,0}(\vec\chi_1\,)\psi_{0,1}(\vec\chi_2\,)\psi_{1,0}(\vec\chi_3\,)$, \\
\phantom{} & $\psi_{0,1}(\vec\chi_1\,)\psi_{0,0}(\vec\chi_2\,)\psi_{1,0}(\vec\chi_3\,)$ \\
\end{tabular}
\end{ruledtabular}
\end{table}

\section{Cross terms between different configurations from central potential}\label{sec:AP1}

\begin{table*}[htb]
\caption{Expectation values of the color-spin operator for the $S=0$ tetraquark state. $<\psi^{k1}_c\psi^{h1}_s|\vec \lambda_i\vec \lambda_j\vec \sigma_i\vec \sigma_j|\psi^{k2}_c\psi^{h2}_s>$}
\label{tab:EHF0}
\resizebox{\textwidth}{!}{%
\begin{tabular}{lcccc}
\hline
\hline
$\vec \lambda_i\vec \lambda_j\vec \sigma_i\vec \sigma_j$ &  $|\psi^c_{\bar 3\otimes 3}\psi^{S=0}_{(0\otimes0)}\rangle$ &  $|\psi^c_{\bar 3\otimes 3}\psi^{S=0}_{(1\otimes1)}\rangle$ & $|\psi^c_{6\otimes \bar 6}\psi^{S=0}_{(0\otimes0)}\rangle$ & $|\psi^c_{6\otimes \bar 6}\psi^{S=0}_{(1\otimes1)}\rangle$ \\
\hline
$|\psi^c_{\bar 3\otimes 3}\psi^{S=0}_{(0\otimes0)}\rangle$ & $\left(8,0,0,0,0,8\right)$&$\left(0,\frac 4{\sqrt 3},-\frac 4{\sqrt 3},-\frac 4{\sqrt 3},\frac 4{\sqrt 3},0\right)$&$\left(0,0,0,0,0,0\right)$&$\left(0,2\sqrt 6,2\sqrt 6,2\sqrt 6,2\sqrt 6,0\right)$\\
$|\psi^c_{\bar 3\otimes 3}\psi^{S=0}_{(1\otimes1)}\rangle$ &$\left(0,\frac 4{\sqrt 3},-\frac 4{\sqrt 3},-\frac 4{\sqrt 3},\frac 4{\sqrt 3},0\right)$&$\left(-\frac83,\frac83,\frac83,\frac83,\frac83,-\frac83\right)$&$\left(0,2\sqrt 6,2\sqrt 6,2\sqrt 6,2\sqrt 6,0\right)$&$\left(0,4\sqrt 2,-4\sqrt 2,-4\sqrt 2,4\sqrt 2,0\right)$\\
$|\psi^c_{6\otimes \bar 6}\psi^{S=0}_{(0\otimes0)}\rangle$ &$\left(0,0,0,0,0,0\right)$&$\left(0,2\sqrt 6,2\sqrt 6,2\sqrt 6,2\sqrt 6,0\right)$& $\left(-4,0,0,0,0,-4\right)$ &$\left(0,\frac {10}{\sqrt 3},-\frac {10}{\sqrt 3},-\frac {10}{\sqrt 3},\frac {10}{\sqrt 3},0\right)$\\
$|\psi^c_{6\otimes \bar 6}\psi^{S=0}_{(1\otimes1)}\rangle$ &$\left(0,2\sqrt 6,2\sqrt 6,2\sqrt 6,2\sqrt 6,0\right)$&$\left(0,4\sqrt 2,-4\sqrt 2,-4\sqrt 2,4\sqrt 2,0\right)$&$\left(0,\frac {10}{\sqrt 3},-\frac {10}{\sqrt 3},-\frac {10}{\sqrt 3},\frac {10}{\sqrt 3},0\right)$&$\left(\frac43,\frac{20}3,\frac{20}3,\frac{20}3,\frac{20}3,\frac43\right)$\\
\hline
\hline
\end{tabular}%
}
\end{table*}

\begin{table*}[htb]
\caption{Expectation values of the color-spin operator for $S=2$ tetraquark state. $<\psi^{k1}_c\psi^{h1}_s|\vec \lambda_i\vec \lambda_j\vec \sigma_i\vec \sigma_j|\psi^{k2}_c\psi^{h2}_s>$}
\label{tab:EHF2}
{
\begin{tabular}{lcccc}
\hline
\hline
$\vec \lambda_i\vec \lambda_j\vec \sigma_i\vec \sigma_j$ &  $|\psi^c_{\bar 3\otimes 3}\psi^{S=2}_{(1\otimes1)}\rangle$ &  $|\psi^c_{6\otimes \bar 6}\psi^{S=2}_{(1\otimes1)}\rangle$ \\
\hline
$|\psi^c_{\bar 3\otimes 3}\psi^{S=2}_{(1\otimes1)}\rangle$ &$\left(-\frac83,\frac43,\frac43,\frac43,\frac43,-\frac83\right)$&$\left(0,4\sqrt 2,-4\sqrt 2,-4\sqrt 2,4\sqrt 2,0\right)$\\
$|\psi^c_{6\otimes \bar 6}\psi^{S=2}_{(1\otimes1)}\rangle$ & $\left(0,4\sqrt 2,-4\sqrt 2,-4\sqrt 2,4\sqrt 2,0\right)$&$\left(\frac43,-\frac{10}3,-\frac{10}3,-\frac{10}3,-\frac{10}3,\frac43\right)$\\
\hline
\hline
\end{tabular}
}
\end{table*}

The color operator $\vec \lambda_i\cdot\vec \lambda_j$ in the Hamiltonian in Eq.~(\ref{eqn::ham}), along with the mass-dependent coefficient in the Cornell-like potential, mixes different color configurations, leading to nonzero cross terms for charmonium-like and bottomonium-like tetraquarks.

Similarly, the color-spin operator $\vec \lambda_i\cdot\vec \lambda_j\vec \sigma_i\cdot\vec \sigma_j$, along with the mass-dependent factor in the spin-spin interaction,
\begin{flalign}\label{AVss}
\vec\lambda_{i}\cdot\vec\lambda_{j}V_{ss}&=\frac{2B_{ij}\sigma_{ij}^3}{3m_im_j\sqrt \pi}e^{-\sigma_{ij}^2r_{ij}^2}\vec\lambda_{i}\cdot\vec\lambda_{j}\vec\sigma_{i}\cdot\vec\sigma_{j}, 
\end{flalign}
mixes different color-spin configurations, also leading to nonzero cross terms for charmonium-like and bottomonium-like tetraquarks.

For reference, the expectation values of $O_{ij}=\vec \lambda_i\cdot\vec \lambda_j\vec \sigma_i\cdot\vec \sigma_j$ for $S=0$ and $S=2$ states are listed in Table~\ref{tab:EHF0} and Table~\ref{tab:EHF2} respectively, with all components given in the format $(O_{12},O_{13},O_{23},O_{14},O_{24},O_{34})$. The mass-dependent factor in Eq.~(\ref{AVss}) induces nonzero cross terms between different color-spin configurations.

In conclusion, the $S=0$ charmonium-like and bottomonium-like tetraquarks are linear combinations of $\psi^c_{\bar 3\otimes3}\psi^{S=0}_{(0\otimes0)}$, $\psi^c_{\bar 3\otimes3}\psi^{S=0}_{(1\otimes1)}$, $\psi^c_{6\otimes\bar 6}\psi^{S=0}_{(0\otimes0)}$, and $\psi^c_{6\otimes\bar 6}\psi^{S=0}_{(1\otimes1)}$. 
Similarly, the $S=2$ charmonium-like and bottomonium-like tetraquarks are linear combinations of $\psi^c_{\bar 3\otimes3}\psi^{S=2}_{(1\otimes1)}$ and $\psi^c_{6\otimes\bar 6}\psi^{S=2}_{(1\otimes1)}$. There is no configuration mixing between S=0 and 2 states, as no cross terms are found.

\bibliography{arXivPwtetra2025}

\end{document}